\documentclass{iopconfser}
\usepackage{eurosym}
\usepackage{amssymb}
\usepackage{amsfonts,cite}
\usepackage{amsmath}
\usepackage{graphicx}
\usepackage{multirow}
\usepackage{epsfig}
\usepackage{xcolor}
\usepackage[utf8]{inputenc}

\begin{document}
	
	\title{A Calogero Model with root string representatives of infinite order Coxeter orbits}
	
	\author{Andreas Fring}
	
	\affil{Department of Mathematics, City St George's, University of London, \\ Northampton Square, London EC1V 0HB, UK}
	
	\email{a.fring@city.ac.uk}

\begin{abstract}
	We present a worked example for the new extensions of the multi-particle Calogero model endowed with infinite Weyl group symmetry of affine and hyperbolic type. Building upon the hyperbolic extension of the $A_3$-Kac-Moody algebra, we construct an explicit realisation of the model in terms of infinite root systems generated from Coxeter orbits. To address the challenge of summing over infinitely many roots, we introduce root string representatives that span the invariant root space while preserving invariance under the affine Weyl group. This approach yields closed-form expressions for the potentials, which by construction are invariant under the full affine Weyl symmetry. Moreover, we demonstrate that in an appropriate infinite-coordinate limit the model reduces smoothly to the conventional four particle $A_3$-Calogero system. Our construction constitutes a systematic method for implementing infinite-dimensional symmetries into Calogero-type models, thus broadening their algebraic and physical applicability.
\end{abstract}

\section{Introduction}

\noindent Recently, several integrable models expressed in term of Lie algebraic roots have been extended to versions incorporating hyperbolic or Lorentzian roots. Consequently, their symmetry with regard to finite Weyl groups has been enlarged to infinite affine, hyperbolic or Lorentzian Weyl group symmetries \cite{fring2019n,fring2021lorentzian,ThesisSam,AFleore,lechtenfeld2022,fring2024toda}. 
Although most of the resulting models, whether of affine Toda scalar field theory or of the multi-particle Calogero–Moser–Sutherland type, are no longer integrable, they display remarkable structural properties and provide a systematic framework for perturbing integrable systems.
 Being of interest in their own right, the study of their underlying mathematics is further motivated by potential applications in modern versions of string theory where such type of extended Kac-Moody algebras play a prominent role \cite{west2000hidden,gaberdiel2002class,damour200210,bossard21}. The mathematical development and conceptual understanding of these type of Kac-Moody algebras, however, remain far from complete, see, for e.g., \cite{KMworkshop}.

 Here, we present a new worked out example of an extended multi-particle Calogero model \cite{Cal1} with hyperbolic symmetry group ${ \bf{g}   } =\mathbf{A}_3^{(1)}$:
\begin{equation}
	H = \frac{1}{2} p^2 +  \sum_{  \alpha \in \Delta_{ \bf{g}   }} \frac{c_{\alpha}}{( \alpha \cdot q )^2 } , \label{calodef}
\end{equation} 
involving generic hyperbolic roots $\alpha \in \mathbb{R}^\ell $, coordinates $q=(q_1,\ldots, q_\ell)$, momenta $p=(p_1,\ldots, p_\ell)$ and coupling constants $c_\alpha \in \mathbb{R}$. The dimension of the phase space, $2 \ell$, is governed by the choice of the representation space of the roots. Crucially the root space $\Delta_{ \bf{g}   }$ is comprised of infinitely many roots. Such type of models were first introduced in \cite{lechtenfeld2022} and can also be formulated for more generic Lorentzian type of algebras \cite{AFleore}.

To be physically meaningful these models pose two main challenges. Firstly, their kinetic terms are of a ghostly nature due to the non-Euclidean metric and thus exhibit the usual difficulties also present in higher time-derivative theories of either having unbounded spectra from below or non-normalisable wave functions. These issues may, however, be overcome (see e.g. recent treatments  \cite{HTDT0,HTDT1,HTDT2}). Secondly, the potential term involves the sum over infinitely many roots. Although this can in principle be made concrete by expanding the roots in terms of simple roots and summing directly, such an approach is cumbersome and somewhat inelegant. Moreover, in general the infinite sums remain at a formal level and can not be carried out explicitly. 

A more systematic approach consists of summing over Coxeter orbits associated with a Coxeter element of infinite order. As a first step closed formulae for the action of arbitrary powers of these elements can then be derived systematically as we will demonstrate explicitly below for the model considered here. The remaining mathematical challenge in the consistent formulation of the models is to identify the suitable representatives of these orbits so that the entire root space is captured. Here, we propose that the simple root representatives should be replaced by infinite root strings. We explicitly carry out this construction for the affine Weyl group of the $\mathbf{A}_3^{(1)}$-case. Furthermore, we demonstrate that, for the associated Calogero model, the infinite sum over each of these representatives can indeed be performed explicitly, thereby yielding a model fully invariant under the infinite affine Weyl group.

Our manuscript is organised as follows: In section 2 we provide the mathematical tools by setting up the entire infinite root space and apply them in section 3 to the extended Calogero model. Our conclusions are stated in section 4.

\section{The infinite root system of the extended $\mathbf{A}_3^{(1)}$-hyperbolic Kac-Moody algebra}
We commence our treatment by constructing the infinite root systems of the affine and hyperbolic versions of $A_3$. The Dynkin diagram for the hyperbolic (or once once-extended affine) Kac-Moody algebra $ \mathbf{A}_3^{(1)}$ and its associated Cartan matrix $K$ are\\
\setlength{\unitlength}{0.58cm} 
\begin{picture}(13.00,5.)(0.0,2.5)
	\thicklines
	\put(1.5,5.0){\Large{ $  \mathbf{A}_3^{(1)}   :$ }}
	\put(5.3,6){{$\small{\alpha_{1}}$}}
	\put(5.3,5){{$\small{\alpha_{2}}$}}
	\put(5.3,4){{$\small{\alpha_{3}}$}}
	\put(6.2,4.2){\line(0,1){0.9}}
	\put(6.2,5.35){\line(0,1){0.9}}
	\put(6.0,5){\Large{$\bullet$}}
	\put(6.0,6){\Large{$\bullet$}}
	\put(6.0,4){\Large{$\bullet$}}
	\put(7.1,5.3){\line(-1,1){0.8}}
	\put(7.1,5.1){\line(-1,-1){0.8}}
	\put(7.0,5){\Large{$\bullet$}}
	\put(7.3,5.2){\line(1,0){0.8}}
	\put(6.8,4.5){{$\small{\alpha_{0}}$}}
	\put(8.0,5){\Large{$\bullet$}}
	\put(7.8,4.5){{$\small{\alpha_{-1}}$}}
	\put(13.0,5){$ \!\!\!  \!\!\! \!\!\! K_{ij}=2 \frac{\alpha_i \cdot \alpha_j}{\alpha_j \cdot \alpha_j}=\left(
		\begin{array}{ccccc}
			 2 & -1 & 0 & 0 & 0 \\
			 -1 & 2 & -1 & 0 & -1 \\
			 0 & -1 & 2 & -1 & 0 \\
			 0 & 0 & -1 & 2 & -1 \\
			 0 & -1 & 0 & -1 & 2 \\
		\end{array}
		\right)_{ij}, $ }
\end{picture}\\
with $ i,j=-1,0,1,2,3$. We represent the hyperbolic $\mathbf{A}_3^{(1)}$-roots in a 4+2 dimensional vector space with Lorentzian inner product 
\begin{equation}
	x\cdot y = x_1 y_1 + x_2 y_2 + x_3 y_3 + x_4 y_4 - x_5 y_6 - x_6 y_5,   \label{innprod}
\end{equation}
for arbitrary 4+2 dimensional vectors $x=(x_1, \ldots x_6)$, $y=(y_1, \ldots y_6)$, as introduced in \cite{gaberdiel2002class}. Then the Cartan matrix $K$ is reproduced with the simple roots
\begin{eqnarray}
	\alpha _{1} \!\! \!\!&=& \!\!\!\!\left( 1,-1,0,0;0,0 \right)
	,\quad \, \,\,\, \,\,\alpha _{2}=\left(0,1,-1,0;0,0 \right) , \quad
    \,\,\,\,  \,\,\alpha _{3}=\left(0,0,1,-1,0;0,0 \right),  \notag	\\
	\alpha _{0} \!\! \!\! &=& \!\! \!\! \left(-1, 0, 0, 1; 1, 0 \right) , \quad \,\,
	\alpha _{-1} =\left(0, 0,0,0;-1,1 \right) .  \notag
\end{eqnarray}
Demanding the length of a generic root in the $\mathbf{A}_3^{(1)}$-root space
\begin{equation}
	\alpha =  q \alpha_{-1} + r \alpha_0 + l \alpha_1 + m \alpha_2 +n\alpha_3 , \qquad   q,r,l,m,n \in \mathbb{Z},      \label{alpha}
\end{equation}
to be 2, leads to the following Diophantine equation
\begin{equation}
l^2-l m-l r+m^2-m n+n^2-n r+q^2-q r+r^2 = 1 \quad \Leftrightarrow \quad \alpha \cdot \alpha =2 . \label{Dio2}
\end{equation}
Thus the entire root space $ \Delta_{ \bf{g}   } $ is generated by listing all possible expressions for the roots  in (\ref{alpha}) subject to the constraint (\ref{Dio2}). While this is in principle a well-defined procedure, it is evidently very involved, somewhat inelegant and in general not explicitly computable in a closed form. A more structured way to proceed is to generate the root system from Weyl reflections or better Coxeter orbits.

In order to set this up this procedure we associate to each of the simple roots $\alpha_i$ a Weyl reflection defined as 
\begin{equation}
	\sigma_i(x) := x- (\alpha_i \cdot x ) \alpha_i, \qquad x \in  \mathbb{R}^\ell, \, \, i=-1,\ldots,3 .
\end{equation}
The symmetries of the Diophantine equation (\ref{Dio2}) are then generated by the Weyl reflections
\begin{eqnarray}
	&& \sigma_{-1}(\alpha) :q \rightarrow  r-q , \quad \,\,\,\,
	\sigma_{0}(\alpha) : r \rightarrow l+n+q-r , \quad  \label{Weylr12} \\
	&&\sigma_{1}(\alpha) : l \rightarrow  m+r-l , \quad 
	\sigma_{2}(\alpha) : m \rightarrow l+n-m  , \quad
	\sigma_{3}(\alpha) : n \rightarrow m+r-n. \notag
\end{eqnarray}
Thus, similarly as in the finite case, we  may start from a simple root, as a trivial solution of (\ref{Dio2}) and act on it consecutively with arbitrary Weyl reflections in any order. As already in the finite case, clearly this is also very involved and cumbersome. A better and controlled process is to act with combinations of Weyl reflections collected in form of a Coxeter element.  Thus, we define the Coxeter element build from Weyl reflections of the affine $\mathbf{A}_3^{(0)}$ algebra
\begin{equation}
	\sigma := \sigma_2 \sigma_0 \sigma_1 \sigma_3,    \label{Coxelaff}
\end{equation}  
and calculate its recursive action on an arbitrary $\mathbf{A}_3^{(1)}$-root of the general form stated in (\ref{alpha})
\begin{equation}
	\sigma^k (\alpha) = \sum_{\nu=-1}^{3} \sum_{\mu=q,r,l,m,n}  a_{\nu\mu}(k) \mu \alpha_\nu, \qquad    a_{\nu\mu}(k) =\left( M^k \right)_{\nu \mu},  \qquad   M= \left(
	\begin{array}{ccccc}
		1 & 0 & 0 & 0 & 0 \\
		1 & 1 & -1 & 2 & -1 \\
		0 & 1 & -1 & 1 & 0 \\
		0 & 2 & -1 & 1 & -1 \\
		0 & 1 & 0 & 1 & -1 
	\end{array}
	\right) .  \label{Cox12}
\end{equation}
We determine the coefficients $ a_{\nu\mu}(k)$ using the general method to compute arbitrary powers of the matrix $M$ as outlined for instance in \cite{epp2010discrete} and applied in \cite{AFleore}. We verify that they satisfy the general recurrence relation
\begin{equation}
a_{\nu\mu}(k+1) = c_1 a_{\nu\mu}(k) + c_2 a_{\nu\mu}(k-1) + c_3 a_{\nu\mu}(k-2)+c_4 a_{\nu\mu}(k-3) .
\end{equation}	
With the matrix $M$ defined in (\ref{Cox12}), we find the same recurrence relation as previously obtained for $A_2^{(0)}$ in \cite{AFleore}
\begin{equation}
	a_{\nu\mu}(k+1) = 2 a_{\nu\mu}(k) -  2 a_{\nu\mu}(k-2) + a_{\nu\mu}(k-3) .
\end{equation}	
Assuming $ a_{\nu\mu}(k)  \sim x^k$ we obtain the same characteristic equation $x^4-2x^3+2x -1 =0$, which is solved by $\lambda_1=-1$, $\lambda_{2,3,4}=1$ so that the general solution  becomes
\begin{equation}
	a_{\nu\mu}(k) = A_{\nu\mu} (-1)^k +  B_{\nu\mu}  + C_{\nu\mu} k+ D_{\nu\mu} k^2 . \label{ABCD}
\end{equation}	
For the appropriate $A_3^{(0)}$ initial conditions, the matrices $A,B,C,D$ are then determined by solving the system of coupled matrix equations
\begin{eqnarray}
	I  \!\! &=&\!\!  a(0) = A + B ,\\
	M \!\! &=& \!\!  a(1) = -A + B + C + D, \\ 
	M^2 \!\!  &=& \!\!  a(2) = A + B + 2 C + 4 D, \\ 
	M^3  \!\! &=& \!\!  a(3) = -A + B + 3 C + 9 D  .
\end{eqnarray}
Using the obtained solutions in (\ref{ABCD}), we find with (\ref{Cox12}) a closed formula for the infinite orbits of the Coxeter element in (\ref{Coxelaff})
\begin{eqnarray}
	\sigma^k (\alpha) &=& q \alpha_{-1} +  \frac{1}{4} \left[    2 r+2 m+q (2 r-2 m-q) (-1)^k+4 (r-l+m-n+r) k+2 q k^2   \right]   \alpha_0   \label{sigmacl}  \\
	&&  +  \frac{1}{2}   \left[    l+n+(l-n) (-1)^k+(2 r-2 l+2 m-2 n-q) k+q k^2 \right]   \alpha_1    \notag \\
		&&  +\frac{1}{4}    \left[  2 m-q+2 r+(2 m+q-2 r) (-1)^k+4 (r-l+m-n) k+2 q k^2  \right]   \alpha_2   \notag   \\
			&&  +\frac{1}{2}  \left[    l+n+(-l+n) (-1)^k+(2 r-2 l+2 m-2 n-q) k+q k^2   \right]   \alpha_3  .  \notag 
\end{eqnarray}  
We can now attempt to construct the entire root space by trying to find the appropriate representatives of the orbits generated by the consecutive action of the Coxeter element. While in the finite case they are easily defined as simple roots dressed appropriately with $\pm$ signs associated to the bi-colouration of the Dynkin diagrams, 	\cite{carter1989simple,dorey1991root,Mass2}, it is not known in general how to proceed in the infinite case. For the stated purpose we define here the following infinite root string representatives
\begin{eqnarray}
	\gamma_0(k) &=& \sigma^k(\alpha_0) =  p_{++}^k \alpha_0 + k \alpha_1 + p_{-+}^k \alpha_2 + k \alpha_3 ,  \label{rootrep0}  \\
	\gamma_1(k) &=& \sigma^k(\alpha_1) =   -k \alpha_0 + p_{+-}^k  \alpha_1 - k \alpha_2 + p_{--}^k \alpha_3  ,  \\
	\gamma_2(k) &=& \sigma^k(\alpha_2) =   p_{-+}^k \alpha_0 + k \alpha_1 + p_{++}^k \alpha_2 + k \alpha_3   ,  \\
	\gamma_3(k) &=& \sigma^k(\alpha_3) =    -k \alpha_0 + p_{--}^k  \alpha_1 - k \alpha_2 + p_{+-}^k \alpha_3  ,   \\
	\gamma_4(k) &=& \sigma_2\sigma^k(\alpha_1) =    -k \alpha_0 + p_{+-}^k  \alpha_1 + (1 - k) \alpha_2 + p_{--}^k \alpha_3  ,   \\
	\gamma_5(k) &=& \sigma_2\sigma^k(\alpha_3) =  -k \alpha_0 + p_{--}^k  \alpha_1 + (1 - k) \alpha_2 + p_{+-}^k \alpha_3   \label{rootrep5}   , 
\end{eqnarray}	
where $p_{mn}^k= [1 +m (-1)^k + n 2k  ] /2$ with $m,n=\pm 1$. We will show that the entire infinite root space can be spanned as
\begin{equation}
	\Delta \equiv \pm  \gamma_i(k), \qquad   i=0,\ldots, 5, \,\, k \in \mathbb{Z},
\end{equation}
and moreover is invariant under all Weyl reflections of $\mathbf{A}_3^{(0)}$. Notice that for $\gamma_i(k)$, $i=0,\ldots, 3$ the strings are being composed of the expressions obtained by acting with $\sigma^k$ on the simple roots $\alpha_i$, so that the $\alpha_i$ may simply be taken as representatives of the corresponding orbits. This is similar to the finite case with the difference that the orbits here are open and contain infinitely many roots. However, the root strings $\gamma_4(k)$ and $\gamma_5(k)$ do not admit such an interpretation. In that case $\sigma^{-k} \sigma_2 \sigma^k (\alpha_1)$ and $\sigma^{-k} \sigma_2 \sigma^k (\alpha_3)$, respectively, would be the representatives, which are $k$-dependent and can therefore not serve as orbit representatives in the conventional sense. For this reason we introduce root strings for all orbits.

In table \ref{table1} we report the action of the $\mathbf{A}_3^{(0)}$ Weyl group elements on the root strings $\gamma_j(k)$ for odd and even $k$ separately. 

\begin{table}[h]
	\centering  
	\begin{tabular}{ l | |  c |  c | c |  c  |  }
	       &  $\sigma_0$  &  $\sigma_1$  &  $\sigma_2$  &  $\sigma_3$  \\   \hline \hline
	     $ \gamma_0 (2 k) $ &  $  - \gamma_0 (-2 k) $    &  $  - \gamma_4 (2 k+1) $    &  $   \gamma_0 (2 k) $  &  $   -\gamma_5 (2 k+1) $    \\  \hline
	      $ \gamma_0 (2 k+1)  $   & $ \gamma_0 (2 k+1)  $      & $   \gamma_5 (-2 k-1) $   & $   -\gamma_0 (-2 k-1) $   & $   \gamma_4 (-2 k-1) $     \\     \hline
	      
	      $ \gamma_1 (2 k) $ &  $ -\gamma_4 (-2 k+1)  $    &  $  - \gamma_1 (-2 k) $     &  $  \gamma_4 (2 k) $  &  $ \gamma_1 (2 k) $    \\  \hline
	       $ \gamma_1 (2 k+1) $ &  $ -\gamma_4 (-2 k)  $    & $    \gamma_1 (2 k+1) $     &  $  \gamma_4 (2 k+1) $  &  $ -\gamma_1 (-2 k-1) $    \\  \hline
	       
	       $ \gamma_2 (2 k) $ &  $   \gamma_2 (2 k)  $     &  $   \gamma_4 (-2 k) $    &  $  -\gamma_2 (-2 k) $  &  $  \gamma_5 (-2 k) $  \\  \hline
	       $ \gamma_2 (2 k+1) $ &  $   -\gamma_2 (-2 k-1)  $    &  $  - \gamma_5 (2 k+2) $    &  $  \gamma_2 (2 k+1) $  &  $  -\gamma_4 (2 k+2) $   \\  \hline
	       
	       $ \gamma_3 (2 k) $ &  $   -\gamma_5 (-2 k+1)  $    &  $ \gamma_3 (2 k) $   &  $  \gamma_5 (2 k) $   &  $  -\gamma_3 (-2 k) $   \\  \hline
	       $ \gamma_3 (2 k+1) $ &  $   -\gamma_5 (-2 k)  $    &  $ - \gamma_3 (-2 k-1) $    &  $  \gamma_5 (2 k+1) $  &  $  \gamma_3 (2 k+1) $    \\  \hline
	       
	       $ \gamma_4 (2 k) $ &  $   -\gamma_1 (-2 k+1)  $    &  $ \gamma_2 (-2 k) $    &  $  \gamma_1 (2 k) $  & $ - \gamma_2 (2 k-1) $   \\  \hline
	       $ \gamma_4 (2 k+1) $ & $   -\gamma_1 (-2 k)  $     & $ - \gamma_0 (2 k) $    &  $  \gamma_1 (2 k+1) $   &  $  \gamma_0 (-2 k-1) $    \\  \hline
	       
	       $ \gamma_5 (2 k) $ &    $   -\gamma_3 (-2 k+1)  $    &  $ - \gamma_2 (2 k-1) $    & $  \gamma_3 (2 k) $   &  $  \gamma_2 (-2 k) $    \\  \hline
	       $ \gamma_5 (2 k+1) $ & $   -\gamma_3 (-2 k)  $    &  $ \gamma_0 (-2 k-1) $    &  $  \gamma_3 (2 k+1) $   & $  -\gamma_0 (2 k) $ 
	\end{tabular}
	\caption{\label{table1}  $\mathbf{A}_3^{(0)}$ Weyl group invariant root strings forming the root space $\Delta$ with table entries $ \sigma_i [ \gamma_j (2k)  ]  $,  $ \sigma_i [ \gamma_j (2k+1)  ]  $ for $i=0,\ldots,3$, $j=0,\ldots,5$.  }
\end{table}
\noindent For instance, we have
\begin{equation}
	\sigma [\gamma_2(2k) ]= \sigma_{2}\sigma_{0}\sigma_{1}[ \gamma_5(-2k) ]=
	\sigma_{2}\sigma_{0} [ -\gamma_2(-2k-1) ]=
	\sigma_{2} [ \gamma_2(2k+1) ] =  \gamma_2(2k+1) .
\end{equation}
By inspection we verify the symmetry by noting that indeed $\sigma_i (\Delta) = \Delta$ for $i=0,1,2,3$. In other words, the infinite root space can be represented by a finite number of infinite root strings. 

Similarly, we compute the repeated action of the Coxeter element build from the hyperbolic $\mathbf{A}_3^{(1)}$-Weyl reflections
\begin{equation}
	\hat{\sigma} := \sigma_{-1} \sigma_2 \sigma_0 \sigma_1 \sigma_3,    \label{Coxel}
\end{equation}  
which is of the general form
\begin{equation}
	\hat{\sigma}^k (\alpha) \!\!= \!\!\sum_{\nu=-1}^{3} \sum_{\mu=q,r,l,m,n}  \hat{a}_{\nu\mu}(k) \mu \alpha_\nu, \qquad    \hat{a}_{\nu\mu}(k) =\left( \hat{M}^k \right)_{\nu \mu},  \qquad   \hat{M}= \left(
	\begin{array}{ccccc}
		0& 1 & -1 & 2 & -1 \\
		1 & 1 & -1 & 2 & -1 \\
		0 & 1 & -1 & 1 & 0 \\
		0 & 2 & -1 & 1 & -1 \\
		0 & 1 & 0 & 1 & -1 
	\end{array}
	\right) .
\end{equation}
Proceeding as before, we find in this case the characteristic equation 
\begin{equation}
	x^5-3 x^3-3 x^2+1 = 0,
\end{equation}
with roots
\begin{equation}
\lambda_1 = \frac{1}{4} \left(1-\sqrt{17}-i\kappa_- \right), \quad
\lambda_2= \lambda_1^*, \quad
\lambda_\pm=  \frac{1}{4} \left( 1+\sqrt{17}\pm \kappa_+ \right), \quad
\lambda_5 =-1.
\end{equation}
$\kappa_\pm =\sqrt{  2 \sqrt{17}   \pm 2      }$.
We determine the coefficient matrix to
\begin{equation}
  \hat{a}(k) = M_1 \lambda_1^k  +M_1^* \lambda_2^k + M_+   \lambda_+^k + M_-   \lambda_-^k + M_5   \lambda_5^k,
\end{equation}
where 
 \begin{equation}
	M_1= \frac{1}{2 \sqrt{17}} \left(
	\begin{array}{ccccc}
		\frac{1}{2} \left(3+\sqrt{17}+i \kappa_-\right) & -2-i \kappa_+ & 1+i \kappa_2 & -2+i \kappa_- & 1+i \kappa _2 \\
		i \kappa_- & \kappa _1 & 2 i \kappa _2 & -2-2 i \kappa _2 & 2 i \kappa _2 \\
		1-i \kappa _2 & -1-i \kappa _2 & \kappa _3 & i \kappa _4 & \kappa _3 \\
		-i \kappa_+ & -2+i \kappa_- & -i \kappa _4 & \frac{\kappa_+^2}{4}+i \kappa _4 & -i \kappa _4 \\
		1-i \kappa _2 & -1-i \kappa _2 & \kappa _3 & i \kappa _4 & \kappa _3 \\
	\end{array}
	\right),
\end{equation}
with  $\kappa_1 = \frac{1}{2} \left(-1+\sqrt{17}-i \sqrt{10 \sqrt{17}-26}\right)$, 
$\kappa_2 = \sqrt{ \sqrt{17} -4   }$, 
$\kappa_3 = \frac{1}{4} \left[-3+\sqrt{17}-i \sqrt{2 \left(25 \sqrt{17}-103\right)}\right]$, $\kappa_4 =\sqrt{  \left( 5 \sqrt{17}-19\right) /2   }$, and

\begin{equation}
	M_\mp =  \frac{1}{2 \sqrt{17}}\left(
	\begin{array}{ccccc}
		2 \lambda _{\mp }-2 & 2\mp \kappa _- & -1\pm \tau _5 & 2\mp \kappa _+ & -1\pm \tau _5 \\
		\mp \kappa _+ & (\tau _3)_{\mp } & \pm 2 \tau _5 & 2\mp 2 \tau _5 & \pm 2 \tau _5 \\
		-1\mp \tau _5 & 1\mp \tau _5 & (\tau _1)_{\pm } & \mp \tau _4 & (\tau _1)_{\pm } \\
		\mp \kappa _- & 2\mp \kappa _+ & \pm \tau _4 & (\tau _2)_{\mp } & \pm \tau _4 \\
		-1\mp \tau _5 & 1\mp \tau _5 & (\tau _1)_{\pm } & \mp \tau _4 & (\tau _1)_{\pm } \\
	\end{array}
	\right),
\end{equation}
with $(\tau_1)_\pm = \frac{1}{4} \left[3+\left(\sqrt{17}\pm \sqrt{206+50 \sqrt{17}}\right)\right]$,
 $(\tau_2)_\pm = \frac{1}{2} \left[-1+\left(\sqrt{17}\pm \sqrt{38+10 \sqrt{17}}\right)\right]$,  \\
  $(\tau_3)_\pm =   \frac{1}{2} \left(1+\left(\sqrt{17}\pm \sqrt{26+10 \sqrt{17}}\right)\right) $,
  $\tau_4 = \sqrt{\frac{1}{2} \left(19+5 \sqrt{17}\right)}$, $\tau_5 =\sqrt{4+\sqrt{17}}$,

\begin{equation}
	M_5= \frac{1}{2} \left(
	\begin{array}{ccccc}
		0 & 0 & 0 & 0 & 0 \\
		0 & 0 & 0 & 0 & 0 \\
		0 & 0 & 1 & 0 & -1 \\
		0 & 0 & 0 & 0 & 0 \\
		0 & 0 & -1 & 0 & 1 \\
	\end{array}
	\right).
\end{equation}
Thus, a closed formula for the infinite orbits of the Coxeter element in (\ref{Coxel}) can be expressed as
\begin{equation}
	\hat{\sigma}^k (\alpha) = (\alpha_{-1},\alpha_0,\alpha_1,\alpha_2,\alpha_3) \cdot (M_1 \lambda_1^k  +M_1^* \lambda_2^k + M_+   \lambda_+^k + M_-   \lambda_-^k + M_5  ) \cdot (q,r,l,m,n)^\intercal .
\end{equation}
We convince ourselves that despite the occurrence of the complicated double square roots the coefficients in front of the simple roots are indeed integers. For instance, we have
{\small 
\begin{eqnarray}
	     \hat{\sigma}^{-2} (\alpha)  \!\!\!\!\!  &=&  \!\!\!\!\!    (l+n-r) \beta _{-1}+(3 l-2 m+3 n-q-r) \beta _0+(4 l-2 m+3 n-2 q-r) \beta _1+(2 l-m+2 n-2 q) \beta _2 \notag \\
	     && +(3 l-2 (m-2 n+q)-r) \beta _3 ,  \notag \\
		\hat{\sigma}^{-1} (\alpha)  \!\!\!\!\!  &=&  \!\!\!\!\! (r-q) \alpha _{-1}+(l+n-q) \alpha _0+(l-m+2 n-q) \alpha_1+(l-m+n) \alpha_2+(2 l-m+n-q) \alpha _3,  \notag\\ 
	\hat{\sigma} (\alpha) \!\!\!\!\! &=&  \!\!\!\!\! (r-l+2 m-n) \alpha _{-1}+(q+r-l+2 m-n) \alpha _0+(r-l+m) \alpha _1+(2r-l+m-n) \alpha _2+(r+m-n) \alpha _3  , \notag \\
	\hat{\sigma}^2 (\alpha) \!\!\!\!\! &=&  \!\!\!\!\!  (q+3 r-2 l+2 m-2 n) \alpha _{-1}+(q+4 r-3 l+4 m-3 n+) \alpha _0+(q+2 r-l+2 m-2 n) \alpha _1 \notag\\
	&&+(2 q+2 r-2 l+3 m-2 n) \alpha _2+(q+2 r-2 l+2 m-n) \alpha _3 .\notag
\end{eqnarray}}
The identification of the root string orbits of $\hat{\sigma}$ is postponed to future work at this stage.

\section{$\mathbf{A}_3^{(1)}$ Weyl group invariant  Calogero model}	

We reformulate now the $\mathbf{A}_3^{(1)}$-extended $\mathbf{A}_3^{(0)}$ invariant Calogero Hamiltonian from (\ref{calodef}) as 

\begin{equation}
	H = \frac{1}{2} p^2 +  	 \sum_{i=0}^{5}  V_i  , \qquad  \text{where} \quad	V_i :=\sum_{n=-\infty}^{\infty}  \frac{ g}{\left[\gamma_i(n) \cdot q \right]^2 }, \label{calonew}
\end{equation} 
For simplicity we have set all coupling constants to be the same, $c_\alpha =g$. The $\gamma_i(n) $ are the root string representatives introduced in (\ref{rootrep0})-(\ref{rootrep5}). We start the discussion of the invariance with the kinetic energy term for which we also need to use the inner product as defined in equation (\ref{innprod}), so that it reads
\begin{equation}
	H_{\text{kin}} = \frac{1}{2} p \cdot p = \frac{1}{2} \left(  p_1^2 + p_2^2 +p_3^2 +p_4^2- 2 p_5 p_6  \right)  . \label{ekin}
\end{equation}
Since Weyl reflections are orthogonal transformation, i.e. $p \cdot p = \sigma_i(p) \cdot \sigma_i(p)$, the kinetic term is invariant by construction under the actions of the entire Weyl group.  More explicitly, we may also realise the transformations on the roots in the dual coordinate space since the Weyl reflections are orthogonal transformation, $\left( \sigma_i (\alpha) \cdot q \right) = \left( \sigma_i^2 (\alpha) \cdot \sigma_i(q) \right) = \left( \alpha\cdot \sigma_i (q) \right) $. The symmetries (\ref{Weylr12}) for our $\mathbf{A}_3^{(1)}$-Weyl group in the six dimensional representation as the action of the Weyl reflections on the roots may then also be realised in the coordinate space as
\begin{eqnarray}
	\sigma_{0}(q) &:& q_1 \rightarrow -q_2-q_3+q_4-q_6 ,  \quad q_2 \rightarrow -q_3 \quad  q_3 \rightarrow -q_2,
	  \quad  q_4 \rightarrow q_1 -q_2-q_3+q_6 ,   \label{sigcoord} \\
	  &&    q_5 \rightarrow q_1 -q_4+q_5+q_6,   \notag \\ 
	\sigma_{1}(q) &:& q_1 \rightarrow q_2 ,  \quad q_2 \rightarrow q_1,  \quad
	\sigma_{2}(q) : q_2 \rightarrow q_3 ,  \quad q_3 \rightarrow q_2, \quad
	\sigma_{3}(q) : q_3 \rightarrow q_4 ,  \quad q_4 \rightarrow q_3.  \notag 
\end{eqnarray}
The affine Coxeter element therefore acts on the coordinates as
\begin{eqnarray}
	\sigma(q)&:& q_1 \rightarrow -q_2,  \quad q_2 \rightarrow -q_2-q_3+q_4-q_6 \quad  q_3 \rightarrow q_1 -q_2-q_3+ q_6 , \quad q_4 \rightarrow -q_3, \quad   \\
	&&  q_5 \rightarrow q_1-q_4 + q_5 + q_6, \quad q_6  \rightarrow q_6 .  \notag
\end{eqnarray}

The action on the momenta is taken to be the same. Thus, using (\ref{sigcoord}) with $q \rightarrow p$, we may now verify explicitly the invariance of the kinetic energy term (\ref{ekin}) under the action of the $\mathbf{A}_3^{(1)}$-Weyl group. The occurrence of linear terms in the momenta in (\ref{ekin}) indicate the aforementioned ghostly nature of the model, an issue that we will not address here. 

The generalisation of the model in the variant in (\ref{calodef}) may be written more explicitly as  
 \begin{equation}
	V(q) =   \!\!\!\!\!\!\!  \sum_{\substack{q,r,l,m,n=0 \\ \text{Diophantine equn}} }^\infty  \!\!\!\!\!   \frac{g}{[ (q \alpha_{-1}+ r \alpha_{0} + l \alpha_{1} + m \alpha_{2} + n \alpha_{3} )   \cdot q]^2} . \label{summ2} 
\end{equation}
 When restricting to particular levels of the Lie algebraic representation, as suggested in \cite{lechtenfeld2022} for a hyperbolic case, one may achieve to carry out some, or possibly all of the infinite sums, at that level. However, the potentials obtained in this manner are not invariant under the infinite Weyl group. In contrast, aiming to achieve the latter we discuss here the version in equation (\ref{calonew}).

For each of the potentials $V_i$ we may in fact compute the infinite sum as we demonstrate in detail for one example. Computing for this purpose the inner product $\gamma_2(n) \cdot q$ gives
\begin{equation}
	V_2(q) = \sum_{n=-\infty}^{\infty} \frac{4 g}{\left[   -{q_1}+{q_2}-{q_3}+{q_4}-{q_6}-2 n {q_6}  + (-1)^n ({q_1}+{q_2}-{q_3}-{q_4}+{q_6})   \right]^2}. 
\end{equation}
Splitting the sum into its even and odd part, i.e. $\sum_n=\sum_{2n} + \sum_{2n-1}$, they are easily evaluated separately when using the general formula $ \sum_{n=-\infty}^{\infty} (A+B n)^{-2} = \pi^2/ [\sin^2(A \pi /B) B^2]    $. In this way we obtain
\begin{equation}
	V_2(q) = \frac{\pi^2 g}{4 q_6^2} \left\{  \frac{1}{\sin^2\left[ \frac{\pi}{2 q_6} (q_2- q_3)  \right]   }  
	+  \frac{1}{\sin^2\left[ \frac{\pi}{2 q_6} (q_1- q_4)  \right]   }  \label{pot111}
	 \right\}  .
\end{equation} 
Calculating the other terms in the same manner, we re-express the potential as
\begin{equation}
	V(q) = \frac{\pi^2 g}{4 q_6^2} \left( V_{12}+ V_{13}+  V_{14} + V_{23} + V_{24} + V_{34}    \right) , \label{affinvpot}
\end{equation} 
where we abbreviated
\begin{equation}
	V_{ij}:= V_{ij}^s+V_{ij}^c, \quad     V_{ij}^s :=  \frac{1}{ \sin^2 \left[ \frac{\pi}{2 q_6} (q_i- q_j)  \right] }, \quad
	 V_{ij}^c  := \frac{1}{ \cos^2 \left[ \frac{\pi}{2 q_6} (q_i- q_j)  \right] }  \quad   i,j= 1,2,3,4.	  
\end{equation}
Thus, the original rational potential has acquired a functional form that is more like a potential of Calogero-Moser-Sytherland type  \cite{Cal1,Mo,Suth3}. We verify explicitly that it is indeed invariant under the affine Weyl group. Indicating how each term transforms under the action of the four generators we have
\begin{eqnarray}
	\sigma_0: &&  V_{12}^s \rightarrow  V_{24}^c  \quad V_{13}^s \rightarrow  V_{34}^c \quad V_{14}^s \rightarrow  V_{14}^s \quad V_{23}^s \rightarrow  V_{23}^s \quad V_{24}^s \rightarrow  V_{12}^c   \quad V_{34}^s \rightarrow  V_{13}^c  ,   \notag \\
	&&  V_{12}^c \rightarrow  V_{24}^s  \quad V_{13}^c \rightarrow  V_{34}^s \quad V_{14}^c \rightarrow  V_{14}^c \quad V_{23}^c \rightarrow  V_{23}^c  \quad V_{24}^c \rightarrow  V_{12}^s   \quad V_{34}^c \rightarrow  V_{13}^s  ,   \notag \\
	\sigma_1: &&  V_{12}^s \rightarrow  V_{12}^s  \quad V_{13}^s \rightarrow  V_{23}^s \quad V_{14}^s \rightarrow  V_{24}^s \quad V_{23}^s \rightarrow  V_{13}^s  \quad V_{24}^s \rightarrow  V_{14}^s   \quad V_{34}^s \rightarrow  V_{34}^s   ,  \notag \\
	&&  V_{12}^c \rightarrow  V_{12}^c  \quad V_{13}^c \rightarrow  V_{23}^c \quad V_{14}^c \rightarrow  V_{24}^c \quad V_{23}^c \rightarrow  V_{13}^c  \quad V_{24}^c \rightarrow  V_{14}^c   \quad V_{34}^c \rightarrow  V_{34}^c ,    \notag \\
	\sigma_2: &&  V_{12}^s \rightarrow  V_{13}^s  \quad V_{13}^s \rightarrow  V_{12}^s \quad V_{14}^s \rightarrow  V_{14}^s \quad V_{23}^s \rightarrow  V_{23}^s  \quad V_{24}^s \rightarrow  V_{34}^s   \quad V_{34}^s \rightarrow  V_{24}^s   ,  \notag \\
	&&  V_{12}^c \rightarrow  V_{13}^c  \quad V_{13}^c \rightarrow  V_{12}^c \quad V_{14}^c \rightarrow  V_{14}^c \quad V_{23}^c \rightarrow  V_{23}^c  \quad V_{24}^c \rightarrow  V_{34}^c   \quad V_{34}^c \rightarrow  V_{24}^c     \notag\\
	\sigma_3: &&  V_{12}^s \rightarrow  V_{12}^c  \quad V_{13}^s \rightarrow  V_{14}^c \quad V_{14}^s \rightarrow  V_{13}^c \quad V_{23}^s \rightarrow  V_{24}^c  \quad V_{24}^s \rightarrow  V_{23}^c   \quad V_{34}^s \rightarrow  V_{34}^c  ,   \notag \\
	&&  V_{12}^c \rightarrow  V_{12}^s  \quad V_{13}^c \rightarrow  V_{14}^s \quad V_{14}^c \rightarrow  V_{13}^s \quad V_{23}^c \rightarrow  V_{24}^s  \quad V_{24}^c \rightarrow  V_{23}^s   \quad V_{34}^c \rightarrow  V_{34}^s   .  \notag
\end{eqnarray}	
The action of the Coxeter element is therefore
\begin{eqnarray}
\sigma: &&  V_{12}^s \rightarrow  V_{34}^c  \quad V_{13}^s \rightarrow  V_{13}^c \quad V_{14}^s \rightarrow  V_{23}^s \quad V_{23}^s \rightarrow  V_{14}^s  \quad V_{24}^s \rightarrow  V_{24}^c   \quad V_{34}^s \rightarrow  V_{12}^c  ,   \notag \\
&&  V_{12}^c \rightarrow  V_{34}^s  \quad V_{13}^c \rightarrow  V_{13}^s \quad V_{14}^c \rightarrow  V_{23}^c \quad V_{23}^c \rightarrow  V_{14}^c  \quad V_{24}^c \rightarrow  V_{24}^s   \quad V_{34}^c \rightarrow  V_{12}^s   .  \notag
\end{eqnarray}
By following through the consecutive action of all elements, we observe the group closure condition and therefore the potential (\ref{affinvpot}) is established to be invariant under the entire infinite Weyl group build from the reflections of $\mathbf{A}_3^{(0)}$. Furthermore, we notice that we recover the standard $A_3$-Calogero model in the infinite limits
\begin{equation}
	\lim_{q_6 \rightarrow \pm \infty} V(q) = 2 g \left[  \frac{1}{(x_1-x_2)^2}+\frac{1}{(x_1-x_3)^2}+\frac{1}{(x_1-x_4)^2}
	+\frac{1}{(x_2-x_3)^2}    \right.  
	   \left. +\frac{1}{(x_2-x_4)^2}+\frac{1}{(x_3-x_4)  
			^2}      \right] .
\end{equation}
Computing the first terms in the version (\ref{summ2}) one can verify explicitly that the two versions of the potential coincide to the lowest levels.

\section{Conclusions}	

In this work, we have provided a consistent formulation of an extended version of the $A_3^{(1)}$-Calogero model. We provided closed analytic expressions for the repeated action of the Coxeter elements $\sigma$ and $\hat{\sigma}$ build from the Weyl reflections of $A_3^{(0)}$ and $A_3^{(1)}$, respectively. For the former case we identified a finite number of infinite root strings as representatives of the Coxeter orbits of infinite order, which can be used to capture the full root space of the extended $A_3^{(0)}$-Kac–Moody algebra. 

In the application to the extended Coxeter model these versions allowed to evaluate the corresponding infinite sums in the Calogero model explicitly. By construction the resulting Hamiltonian, comprising both the kinetic and potential contributions, was shown to be invariant under the affine Weyl group. The obtained potential was shown to reduce smoothly to the familiar finite $A_3$-Calogero interaction the infinite limit of one coordinate, thereby confirming the consistency of the construction. While the model retains ghost-like features in its kinetic term, due to the Lorentzian structure, we believe these issues can be treated within recently developed ghost-free quantisation frameworks. The verification for this is left for future work dealing with the quantisation of the model.

Our findings open the path to further investigations of integrable models associated with infinite-dimensional Lie algebras and may have applications in the study of extended symmetries in field theory and many-body physics. Naturally, there are a number of open questions to fully understand these models and their underlying mathematical structures. From a mathematical point of view it would be interesting to identify the root string representatives for more examples and ultimately unravel the more generic principles that determine them. Concerning the extended Calogero models the challenge remains to carry out the infinite sums in more cases, especially for the full hyperbolic case and extensions to the Lorentzian cases.  Yet unexplored are also the generalisations of these extended systems to complex $\cal{PT}$-symmetric versions of the Calogero models in analogy to the finite treatment in \cite{Mon1}.

\medskip

\noindent \textbf{Acknowledgments:}  I would like to thank Cestmir Burdik for his continued efforts in organising the International Conference on Integrable Systems and Quantum Symmetries, which will be celebrating its 30th anniversary next year.

\newif\ifabfull\abfulltrue


\begin{thebibliography}{10}
	
	\bibitem{fring2019n}
	A.~Fring and S.~Whittington,
	\newblock n-Extended Lorentzian Kac--Moody algebras,
	\newblock Lett. Math. Phys, 1--22 (2020).
	
	\bibitem{fring2021lorentzian}
	A.~Fring and S.~Whittington,
	\newblock Lorentzian Toda field theories,
	\newblock Rev. in Math. Phys. {\bf 33}(06), 2150017 (2021).
	
	\bibitem{ThesisSam}
	S.~Whittington,
	\newblock {\em Extensions of Integrable Quantum Field Theories Based on
		Lorentzian Kac-Moody Algebras},
	\newblock PhD thesis, City, University of London, 2022.
	
	\bibitem{AFleore}
	F.~Correa, A.~Fring and O.~Quintana, 
	\newblock Infinite affine, hyperbolic and Lorentzian Weyl groups with their associated Calogero models,
	\newblock J. Phys. A: Math. Theor. {\bf 57}, 055203 (2024).
	
	\bibitem{lechtenfeld2022}
	O.~Lechtenfeld and D.~Zagier,
	\newblock A hyperbolic Kac-Moody Calogero model,
	\newblock JHEP {\bf 2024.6} 1-31  (2024).
	
	\bibitem{fring2024toda} 
	A.~Fring,
	\newblock Toda field theories and Calogero models associated to infinite Weyl groups,
	\newblock J. of Phys.: Conf. Series {\bf 2912}, 012021 (2024). 
	
	\bibitem{west2000hidden}
	P.~West,
	\newblock Hidden superconformal symmetry in M-theory,
	\newblock JHEP {\bf 2000}, 007 (2000).
	
	\bibitem{gaberdiel2002class}
	M.~R. Gaberdiel, D.~I. Olive, and P.~C. West,
	\newblock A class of Lorentzian Kac--Moody algebras,
	\newblock Nucl. Phys. B {\bf 645}, 403--437 (2002).
	
	\bibitem{damour200210}
	T.~Damour, M.~Henneaux, and H.~Nicolai,
	\newblock E10 and a small tension expansion of M theory,
	\newblock Phys. Rev. Lett. {\bf 89}(22), 221601 (2002).
	
	\bibitem{bossard21}
	G.~Bossard, A.~Kleinschmidt, and E.~Sezgin,
	\newblock A master exceptional field theory,
	\newblock JHEP {\bf 2021}(6), 1--148 (2021).
	
	
	\bibitem{KMworkshop}
	G.~Bossard, A.~Kleinschmidt, and H.~Nicolai,
	\newblock Mini-Workshop: Infinite-Dimensional Kac-Moody Lie Algebras in Supergravity and M Theory
	\newblock Oberwolfach Reports {\bf 21}(4), 3055--3102 (2025).
	
	
	\bibitem{Cal1}
	F.~Calogero,
	\newblock Ground state of one-dimensional N body system,
	\newblock J. Math. Phys. {\bf 10}, 2197--2200 (1969).
	
	
	\bibitem{HTDT0}
	A.~Fring, T.~Taira and B.~Turner, 
	\newblock Quantisations of exactly solvable ghostly models,
	\newblock J. Phys. A: Math. Theor. {\bf 58}, 2353013 (2025).
	
	\bibitem{HTDT1}
	A.~Felski, A.~Fring and B.~Turner, 
	\newblock Lie symmetries and ghost-free representations of the Pais-Uhlenbeck model,
	\newblock Preprint arXiv:2505.07869 (2025).
	
	\bibitem{HTDT2}
	A.~Fring, T.~Taira and B.~Turner, 
	\newblock Ghost-Free Quantisation of Higher Time-Derivative Theories via Non-Unitary Similarity Transformations,
	\newblock Preprint arXiv:2506.21400 (2025).
	
	\bibitem{epp2010discrete}
	S.~S. Epp,
	\newblock {\em Discrete mathematics with applications},
	\newblock Cengage, Boston, MA, 2020.
	
	
	\bibitem{carter1989simple}
	R.~W. Carter,
	\newblock {\em Simple groups of Lie type}, volume~22,
	\newblock John Wiley \& Sons, 1989.
	
	\bibitem{dorey1991root}
	P.~Dorey,
	\newblock Root systems and purely elastic S-matrices,
	\newblock Nucl. Phys. B {\bf 358}, 654--676 (1991).
	
	\bibitem{Mass2}
	A.~Fring, H.~C. Liao, and D.~Olive,
	\newblock The mass spectrum and coupling in affine Toda theories,
	\newblock Phys. Lett. B {\bf 266}, 82--86 (1991).
	
	\bibitem{Mo}
	J.~Moser,
	\newblock Three integrable Hamiltonian systems connected with isospectral
	deformations,
	\newblock Adv. Math. {\bf 16}, 197--220 (1975).
	
	\bibitem{Suth3}
	B.~Sutherland,
	\newblock Exact results for a quantum many body problem in one- dimension,
	\newblock Phys. Rev. {\bf A4}, 2019--2021 (1971).
	
	\bibitem{Mon1}
	A.~Fring and M.~Smith,
	\newblock Antilinear deformations of Coxeter groups, an application to Calogero
	models,
	\newblock J. Phys. A {\bf 43}, 325201 (2010).
	
\end{thebibliography}
\end{document}